\documentclass[preprint,showpacs,showkeys,aps,prb]{revtex4}
\usepackage[english]{babel}

\usepackage[dvips]{graphicx}
\usepackage{amssymb}

\begin{document}  


\title{Single-Speed Molecular Dynamics of Hard Parallel Squares and Cubes}

\author{Wm. G. Hoover and Carol G. Hoover \\
Ruby Valley Research Institute \\ Highway Contract 60,
Box 598, Ruby Valley 89833, NV USA    \\ }

\author{Marcus N. Bannerman \\
School of Chemical Engineering and Analytical Science \\
The University of Manchester \\
Manchester M60 IQD, United Kingdom \\ }

\date{\today}




\vskip 0.5cm

\begin{abstract}
The fluid and solid equations of state for hard parallel squares and cubes
are reinvestigated here over a wide range of densities.  We use a novel
single-speed version of molecular dynamics.  Our results are
compared with those from earlier simulations, as well as with the predictions
of the virial series, the cell model, and Kirkwood's many-body single-occupancy
model.  The single-occupancy model is applied to give the absolute
entropy of the solid phases just as was done earlier for hard disks and
hard spheres.  The excellent
agreement found here with all relevant previous work shows very clearly
that configurational properties, such as the equation of state, do not
require the maximum-entropy Maxwell-Boltzmann velocity distribution.  For
both hard squares and hard cubes the free-volume theory provides a good
description of the high-density solid-phase pressure.  Hard parallel
squares appear to exhibit a second-order melting transition at a
density of 0.79 relative to close-packing.  Hard parallel cubes have a
more complicated equation of state, with several relatively-gentle
curvature changes, but nothing so abrupt as to indicate a first-order melting
transition.  Because the number-dependence for the cubes is relatively
large the exact nature of the cube transition remains unknown.

\end{abstract}

\maketitle

\section{Introduction}

Hard parallel squares and cubes have undergone extensive
study\cite{b1,b2,b3,b4,b5,b6,b7,b8,b9}.  Most of the hard-particle work
motivating our present efforts is roughly
50 years old:  Monte Carlo simulation\cite{b6} indicated the absence of a
first-order transition for hard parallel squares, while
corresponding molecular dynamics simulations suggested its presence\cite{b5}.
Because computers are now so much faster it is appropriate to reinvestigate
this problem as well as the three-dimensional hard-cube analog.

In addition to the equilibrium equation of state, mixtures, transport
coefficients, and various correlations have all been previously
studied for squares and cubes.  The most basic questions for statistical
mechanics are the existence and nature of the melting transition for
these two simple models.  This question has been thoroughly settled for
hard spheres, which exhibit a first-order transition between two
coexisting phases, fluid and solid\cite{b10}. Despite
hundreds of investigations, following the pioneering work of Alder, Jacobsen,
Wainwright, and Wood\cite{b11,b12}, the evidence is still not complete for
disks, squares, and cubes\cite{b13,b14}.  This uncertainty helped motivate
the present work.

The two-dimensional squares model
and its three-dimensional analog, the hard parallel cube model, are
somewhat more tractable than disks and spheres because the square and cube
potential functions are products of one-dimensional functions,
$$
\phi_{\rm squares} = \phi(|x|)\phi(|y|) \ ; \
\phi_{\rm cubes} = \phi(|x|)\phi(|y|)\phi(|z|) \ ; \
$$
$$
\phi(0<x<1) = \infty \ ; \ \phi(x>1) = 0 \ .
$$
The analytical simplicity due to these factorizations is a major
motivation for the study of
these systems, with an understanding of the melting transition a key goal.
A good deal of the prior work lies twenty years or more in the past, so that
today's enhanced computer speeds can lead to more precise conclusions than
could the earlier work.

Throughout this work we set the mass and distance scales by imagining hard
particles of unit mass and sidelength.  The particles cannot rotate, acting
as if their moments of inertia were infinite. The particles remain 
forever parallel, with their edges lined up with the $x$, $y$, and $z$ axes.
See Figure 1 for a sample two-dimensional fluid configuration. 

\begin{figure}
\includegraphics[height=7cm,width=6cm,angle=-90]{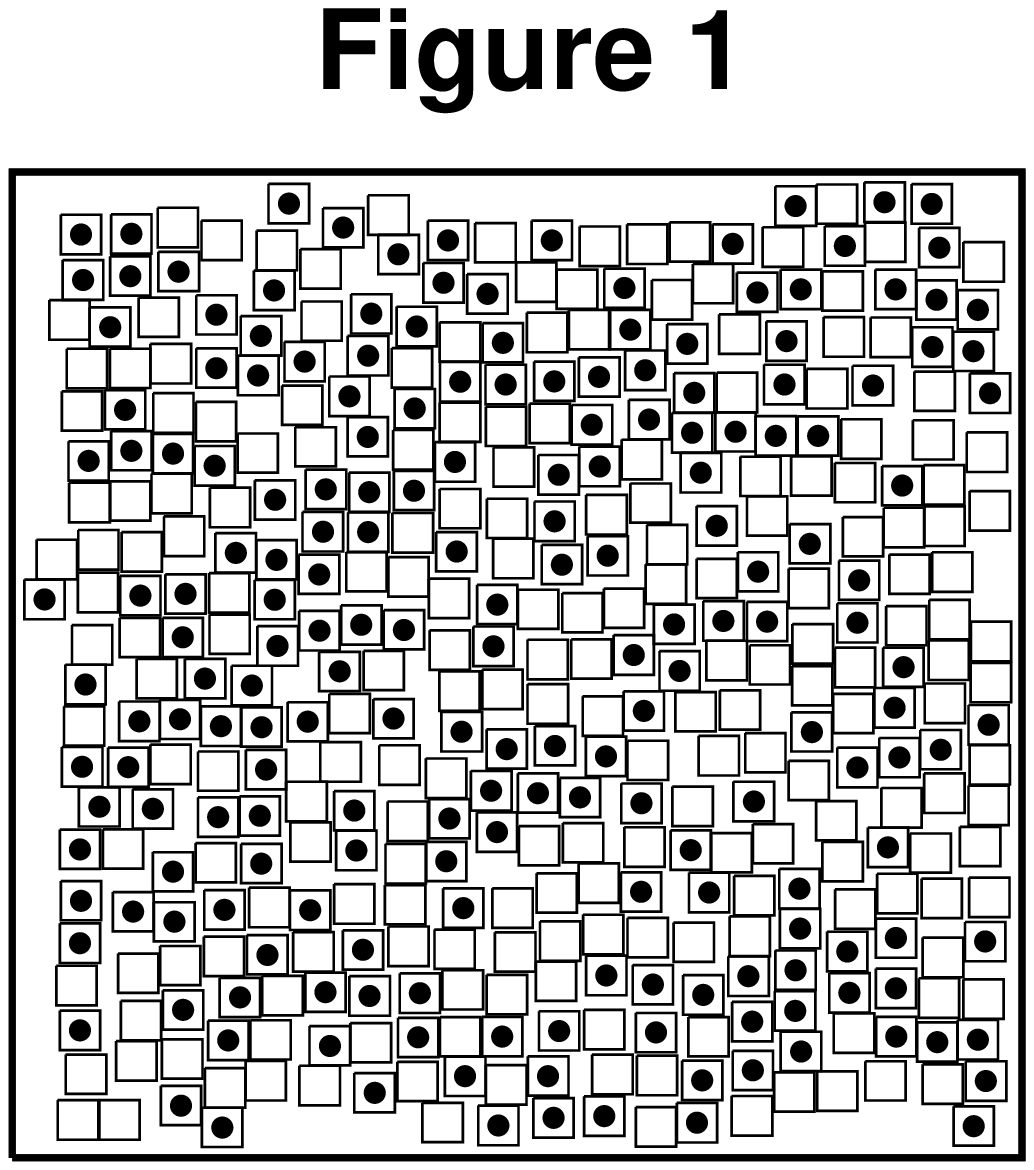}
\vspace{.5in}
\caption{
A sample periodic configuration of $N = 400$ hard parallel squares at two
thirds the close-packed density, $V = 600$.  The figure illustrates a fluid.
In the initial condition the squares with dots occupied the even-numbered
rows of a perfect square lattice.
}
\end{figure}

The parallel square and cube
models simplify the evaluation of the phase integrals derived from Gibbs'
statistical mechanics. Both squares and cubes have fluid and solid
phases, though until now the number-dependence of the dynamics and the
thermodynamics has concealed the exact nature of the fluid-solid
transitions.  Gibbs' statistical mechanics shows that the pressure can be
calculated from the ``configurational integral'' $Q_N(V,T)\cite{b15,b16}$:
$$
Q_N \equiv \int_V dr_1 \dots \int_V dr_Ne^{-\Phi /kT}/N! \ ; \
\Phi = \sum_{i<j}^N\phi_{ij} \ ;
$$
$$
PV/NkT = (\partial \ln Q_N/\partial \ln V)_T \ .
$$
$Q_N$ is the integral over all distinct arrangements of $N$ particles within
a box of volume $V$ at the temperature $T$.  $\Phi $ is the potential
energy, either infinity or
zero for the square and cube models.  In the present work we set the energy
scale by choosing Boltzmann's constant and the temperature equal to unity,
$kT = 1$.

The Mayers carried out an exact low-density series expansion of the
pressure\cite{b15}, the ``virial expansion''.   For squares and cubes the
series' coefficients, the virial coefficients, have been evaluated,
analytically, through
the seventh term\cite{b1,b2,b3}. A convenient extrapolation method for the
series is provided by ratios of polynomials, ``Pad\'e
approximants''\cite{b17,b18}, of the type given in the Appendix.

At high density, where neither the density series nor its extrapolation
are useful, a ``free-volume'' approach, exact near close packing\cite{b19,b20},
can be used.  For $D$-dimensional hard cubes of unit sidelength in a rigid
box of sidelength $L = V^{1/D}$, the configurational integral is
$DN$-dimensional, but easy to approximate using ideas borrowed from Tonks'
one-dimensional work\cite{b21} and the Eyring-Hirschfelder cell
model\cite{b22}.  If for $D=3$ we assume that the cubes are ordered in
$N^{2/3}$
columns parallel to the $z$ axis and allowed to move independently in the
$x$ and $y$ directions, as in the self-consistent cell model of Figure 2,
the configurational integral over the $x$ and $y$ coordinates gives:
$$
 \prod^N (\int dx\int dy) \rightarrow [(V/N)^{1/3} - 1]^{2N} \ .
$$
Because the arbitrary ordering of the particles can be chosen in $N!$
distinct way, this ordering degeneracy exactly compensates for the factor
of $1/N!$ in the definition of $Q$.  The remaining integrals in the $z$
direction give Tonks' result for the one-dimensional hard-rod
configurational integral:
$$
\prod^N (\int dz) \rightarrow
[(V^{1/3} - N^{1/3})^{N^{1/3}}/(N^{1/3})!]^{N^{2/3}} \simeq
$$
$$
\left[ [(V/N)^{1/3} - 1]^{N^{1/3}}e^{N^{1/3}}\right]^{N^{2/3}} =
 \  [(V/N)^{1/3} - 1]^Ne^N \ ,
$$
resulting in the lower bound:
$$
Q_N(V,T) > [(V/N)^{1/3} - 1]^{3N}e^N \ .
$$
For $D$-dimensional hard cubes the ordinary Eyring-Hirschfelder 
cell model exceeds this estimate by a factor of $(2^D/e)^N$. See the
central illustration in Figure 2 for a sketch of this cell model.

\begin{figure}
\includegraphics[height=13cm,width=9cm,angle=-90]{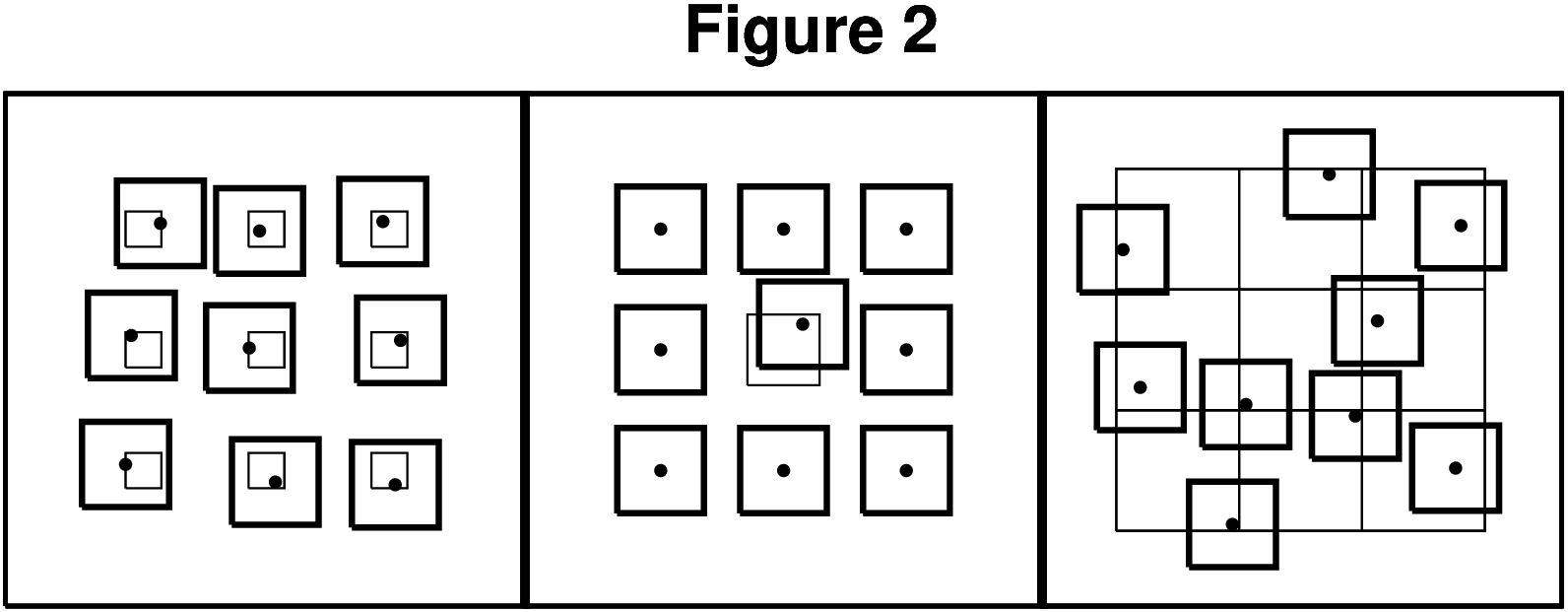}
\caption{
The self-consistent cell model is shown at the left, and allows for the
simultaneous independent motion of the centers (shown as dots) of all
particles within the individual light squares of accessible states.  In
the self-consistent cell models all $N$ particles are treated alike.  The
more nearly accurate Eyring-Hirschfelder cell
model shown in the center, has all the neighboring particles fixed
while the central particle wanders over a much larger ``free volume'',
four times bigger (for squares) than in the self-consistent case for
$\rho > 0.25$.  The single-occupancy system, shown at the right, confines
(the center of) each particle to a square of area $V/N$.  Unlike the
cell models, which reduce to simple one-body problems, the single-occupancy
model is as complex to treat analytically as is the full unconstrained
many-body problem.
}
\end{figure}

The free-volume equation of state results from either approach, the
lower bound or the cell model,
$$
PV/NkT = 1/(1-\rho^{1/D}) \ ; \ \rho > (1/2)^D \ {\rm for} \ D>1 \ .
$$
Our single-speed molecular dynamics results --- see Sections IV and V ---
suggest that this approximation is exact within terms of order unity,
 for hard parallel squares or
cubes near close packing.  For instance, a 128 000-collision simulation
with 1000 hard parallel cubes at a density of 0.95 gave
$PV/NkT = 58.99 \pm 0.02$, equal to the free-volume compressibility
factor, which is also 58.99 at this density.
Our single-speed molecular dynamic results agree perfectly well with earlier
results based on the Maxwell-Boltzmann velocity distribution.

This report is organized as follows.  In Section II the Mayers' virial series
is reviewed for squares and cubes.  Section III describes the
Eyring-Hirschfelder cell model approach to their thermodynamic properties.  The
cell model is specially useful for squares and cubes.  We include here the
details of Kirkwood's many-body single-occupancy model, a nearly exact
description of the solid phase.  Section IV describes the kinetic theory used
to analyze the molecular dynamics simulations.  The simulations and their
results are described in the following Section V. Section VI is devoted to the
nature of the phase transition(s) for squares and cubes, with Section VII a
summary of our results and conclusions, including an attempt to reconcile
our findings with the work of Jagla\cite{b13}, Groh, and Mulder\cite{b14}.

\section{Low Density and the Mayers' Virial Series}

There are plenty of theoretical approaches --- series expansions, cell
models, integral equations --- to the equation of state and
thermodynamic properties.  Only one of them is rigorously correct ---
the Mayers' ``virial expansion'' of pressure as a series in the
density\cite{b15}.  This virial expansion gives a fairly good
representation of the entire fluid equation of state for squares and
for cubes.  The hard-square and hard-cube virial series were carried
out through seven terms in 1960\cite{b2,b3}:
$$
(PV/NkT)_{2D} = 1 + 2\rho + 3\rho ^2 + 3.66667\rho ^3 + 3.72222\rho ^4
+ 3.02500\rho ^5 + 1.65065\rho ^6 + \dots \ .
$$
$$
(PV/NkT)_{3D} = 1 + 4\rho + 9\rho ^2 + 11.33333\rho ^3 + 3.15972\rho ^4
-18.87963\rho ^5 -43.50543\rho ^6 + \dots \ .
$$
See again the Appendix for convenient Pad\'e extrapolations of these
truncated series. The negative $B_6$ and $B_7$ for cubes are notable
as the first known instance
in which hard particles definitely display negative (tensile) contributions to
the virial expansion of the pressure.  It is still unknown whether or
not hard disks and hard spheres have such negative contributions.

In 1960 progress beyond $B_7$ was stalled by limited computer resources. 
The evaluation of $B_7$ required computing 468 separate integrals over
the relative coordinates describing seven particles.  The integrands
are products of from seven to 21 of the Mayers' ``$f$ functions'',
$$
f(r) = e^{-\phi/kT} - 1 \ .
$$
To simplify the integrals' evaluation Ree and Hoover introduced the identity
$$
1 \equiv e^{-\phi/kT} - f \ ,
$$
for all pairs of particles not linked by $f$ functions in the
integrands, leading to a reduced number of integrals and to substantially
better numerical accuracy in Monte Carlo calculations of the
higher $B_n$.  The number of integral types contributing to $B_7$ was
reduced in this way from 468 to 171\cite{b17,b18}.

If, as is the case for hard disks, there were a melting transition for
squares at about four-fifths the close-packed density, $\rho \simeq
0.80$, then the last of these known terms in the series, would make a
contribution of about
five percent to the total melting pressure.  Techniques already
developed for hard disks and spheres\cite{b18} could be applied to generate an
additional three terms in the series.  For $B_8$, $B_9$, and $B_{10}$
2606, 81 564, and 4 980 756 integral types need to be evaluated.

\section{High Density: the Eyring-Hirschfelder and Single Occupancy Models}

At higher density, near close packing, ``cell models'' are useful
approximations.  These models are based on the notion that particles
sweep out a ``free volume'' bounded by their neighbors.  Certain
aspects of this idea are exactly correct\cite{b19,b20}.  This is the
consequence of two facts: first, configurational
properties are mass-independent in classical statistical mechanics;
second, the dynamical evolution of a very light particle, moving rapidly
in the presence of nearly stationary neighbors, {\em does} sweep out a free
volume as time goes on.  It should in fact be possible to derive the Mayers'
virial series by considering this point of view in detail.

A much more complicated, but still cell-like, ``single-occupancy'' model
can be constructed.  This single-occupancy model gives a near-exact
(within terms of order unity in $PV/NkT$) description of the solid phase.
In the single-occupancy
model each particle is constrained to one of $N$ nonoverlapping cells.
Because vacancies and dislocations, as well as excursions outside such
cells, are unimportant to the thermodynamics of the solid phase, the
single-occupancy configurational integral,
$$
Q_{\rm SO} \equiv \int_{(V/N)} dr_1 \dots \int_{(V/N)} dr_Ne^{-\Phi /kT} \ ,
$$ 
gives nearly the same solid-phase pressure-volume equation of state as does the
exact configurational integral $Q_N$.  Notice that the $1/N!$ appearing
in $Q_N$ is absent in $Q_{\rm SO}$.  This is because each particle is
restricted to occupy a particular cell.  By including collisions with
cell walls it is easy to modify a molecular dynamics simulation to
compute single-occupancy properties, as we detail in Section V.

Besides exact free-volume measurements\cite{b20}, there are several approximate
methods for estimating the free volume.  In the self-consistent cell
model, all particles are distributed so near their lattice sites that
no overlaps can occur.  In the alternative inconsistent, but more nearly
accurate, Eyring-Hirschfelder cell model, the motion of a single particle
is considered, with all its neighbors held fixed at their lattice sites.
In either case the approximate partition function includes the $N$th
power of the cell-model free volume:
$$
Z(N,V,T) \equiv v_f^N/\lambda ^{DN} \ ; \ \lambda ^2 = h^2/2\pi mkT \ .
$$
As is usual $h$ is Planck's constant and $\lambda $ is de Broglie's
wavelength.
Both of the cell models and the single-occupancy model are illustrated for
hard parallel squares in Figure 2.

At high density, these forms of the cell model, plus
various approximate bounds on the hard square partition function all
suggest that the ``free volume'' equation of state:
$$
PV/NkT = 1/(1-\rho^{1/D}) \ .
$$
is asymptotically correct near the close-packed limit,
$ \rho \rightarrow 1$.  At a density of $2^{-D}$ with $D>1$ the
Eyring-Hirschfelder cell model allows the
central ``wanderer'' particle to escape its cell.  The free volume changes
there, discontinuously, from an intensive localized volume to a netlike
extensive volume --- the total volume $V$ less the exclusion volumes of
the  $N-1$ particles fixed at their lattice sites.  At this ``percolation
transition''\cite{b9,b20} the model pressure jumps from the free-volume value,
$\rho kT/(1-\rho^{1/D})$, to infinity.

Monte Carlo hard-square simulations showing the absence of a sharp
fluid-solid transition\cite{b5} contradict molecular dynamics
work\cite{b4,b6}, also carried out in the early 1970s.  The molecular
dynamics results suggested a van der Waals loop joining the two phases.
In the present work we measure the equation of state using molecular
dynamics with the special single-speed velocity distribution described
in the next Section. We also use single-occupancy simulation results to
measure the solid-phase entropy directly.

\section{Single-Speed Molecular Dynamics for Squares and Cubes}

The factorization of the partition function into a kinetic part and
a configurational part suggests that {\em any} reasonable velocity
distribution, with vanishing total momentum and capable of reaching all
configurations, can be used for computing configurational properties.
In the present work we choose the $x$ and $y$ and $z$ velocity
components all equal to $\pm1$, corresponding to unit isotropic
temperature:
$$
v_x^2 = v_y^2 = v_z^2 \equiv kT/m = 1 \ .
$$

Parallel hard squares and cubes move and collide as if their moments
of inertia were infinite.  The particles do not rotate when they
collide, but simply exchange $x$ or $y$ or $z$ momenta (in the center
of mass system of coordinates) on collision.  Thus the velocity 
distribution is unchanged by particle collisions.  In single-occupancy
simulations the cell walls change this.  Then the center of each square
or cube is confined to an individual cell of volume $V/N$.  Collisions
at the cell walls simply reflect the $x$ or $y$ or $z$ momentum
perpendicular to the confining wall.  Whenever a particle is
reflected by a cell wall the center-of-mass momentum is shifted, by
$\pm2/N$.

In all of our simulations the number of particles, the density, and the
temperature are fixed.  From the measured all-pairs particle-particle collision
rate we determine the pressure.  In the single-occupancy case note that
the cell walls make no special nonideal contribution to the pressure.
The kinetic part of the pressure is still given by $(PV/NkT)_K = 1$.
We choose to calculate the total pressure directly from the measured
all-pairs collision rate $\Gamma $, using the exact relation:
$$
PV/NkT = (PV/NkT)_K + (PV/NkT)_\Phi = 1 + B_2\rho(\Gamma /\Gamma _0) \ .
$$
The dot product $(F \cdot r)_{ij}$ is the same for every collision:
$$
(F \cdot r)_{ij} \equiv F_{ij}\cdot r_{ ij} \equiv
-\nabla _i\phi _{ij}\cdot (r_i - r_j) = 2kT \ .
$$
The time average, which gives the potential contribution to $PV$, is
computed by summing all the $C$ collisional $ij$ pair contributions
taking place during the sufficiently long time $t$:
$$
(1/t)\sum_C (F \cdot r)_{ij} = (1/t)\sum_C kT \ . 
$$
As a consequence, the ``virial-theorem pressure'' with
single-speed dynamics is identical to the ``collision-rate pressure'':
$$
PV/NkT = 1 + (1/DNkT) \sum_{i<j} \langle (F \cdot r)_{ij} \rangle 
       = 1 + B_2\rho(\Gamma /\Gamma _0) \ .
$$

The low-density collision rate $\Gamma_0$ can be calculated in either
of two different ways, both leading to the same result.  A relatively
complex approach is to calculate separate cross-sections and collision
probabilities for relative
speeds of $(\pm \sqrt{4},\pm \sqrt{8},\pm \sqrt{12})$ (for cubes).  The 
simpler approach multiplies the probability for a collision of cubes
$i$ and $j$ in the $x$ direction by 3 and by $N(N-1)/2$, the number of
pairs of particles, giving:
$$
(\Gamma_0/N) = 2\rho \  ({\rm squares}) \ ; 
$$
$$
(\Gamma_0/N) = 6\rho \  ({\rm cubes}) \ .
$$
To confirm these simple relations and to check that the single-speed
dynamics gives the same pressure as does Maxwell-Boltzmann dynamics, we
measured the collision rate for 1000 cubes at a density of $0.1$ for a
run with 512 000 collisions.  The collision rate per particle (collisions
per unit time divided by the total number of particles) was 0.751669,
giving a compressibility factor of 
$$
PV/NkT = 1 + 0.1\times4\times(0.751669/0.6) = 1.5011 \pm 0.0001 \ ,
$$
in excellent agreement with van Swol and Woodcock's 1987
calculation\cite{b9}, $1.5016 \pm 0.004$.
We must stress that the simple velocity
distribution $(\pm1,\pm1,\pm1)$, because the system is configurationally
ergodic, gives the same pressure as would a Maxwell-Boltzmann
distribution (though with considerably less effort).

Both the low-density and high-density regions are well understood for squares
and cubes.  Our main interest is in the square and cube analogs of what Wood
aptly called ``the region of confusion'' for hard disks, where the fluid and
solid phases come and go, but with a pace so slow that meaningful averages
are hard to obtain.  We emphasize the region of confusion in the following
two Sections, which are devoted to the results of our simulations. Our
single-occupancy results, together with thermodynamic integration,
$$
d(S/Nk)_T = -(PV/NkT)d\ln\rho \ ,
$$
make it possible to determine the relative stabilities of the fluid and
solid phases as functions of density.

\section{Pressure and Entropy from Single-Speed Molecular Dynamics}

\begin{figure}
\includegraphics[height=9cm,width=9cm,angle=-90]{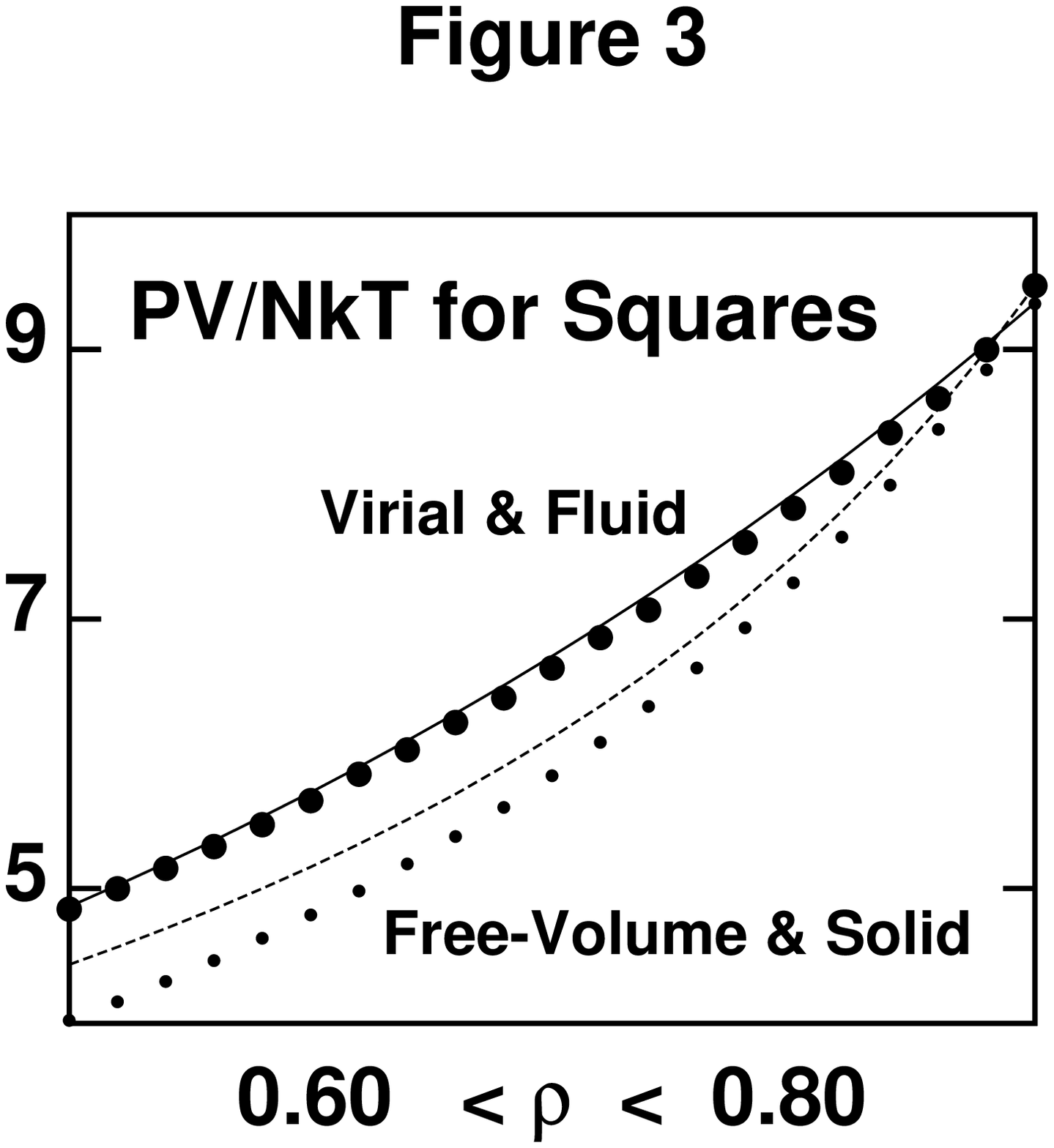}

\caption{
Compressibility factor for 400 hard parallel squares (dots) compared
with the predictions (lines) of the 7-term virial series and the
free-volume theory.  The upper set of larger dots represents unconstrained
molecular dynamics while the lower set of smaller dots represents
single-occupancy simulations.  The fluid points correspond to 400 000
collisions each; the solid points correspond to 4 000 000 collisions each.
The dots shown represent 21 simulations, equally spaced in density from
0.60 to 0.80, inclusive.
}
\end{figure}

\begin{figure}
\includegraphics[height=9cm,width=9cm,angle=-90]{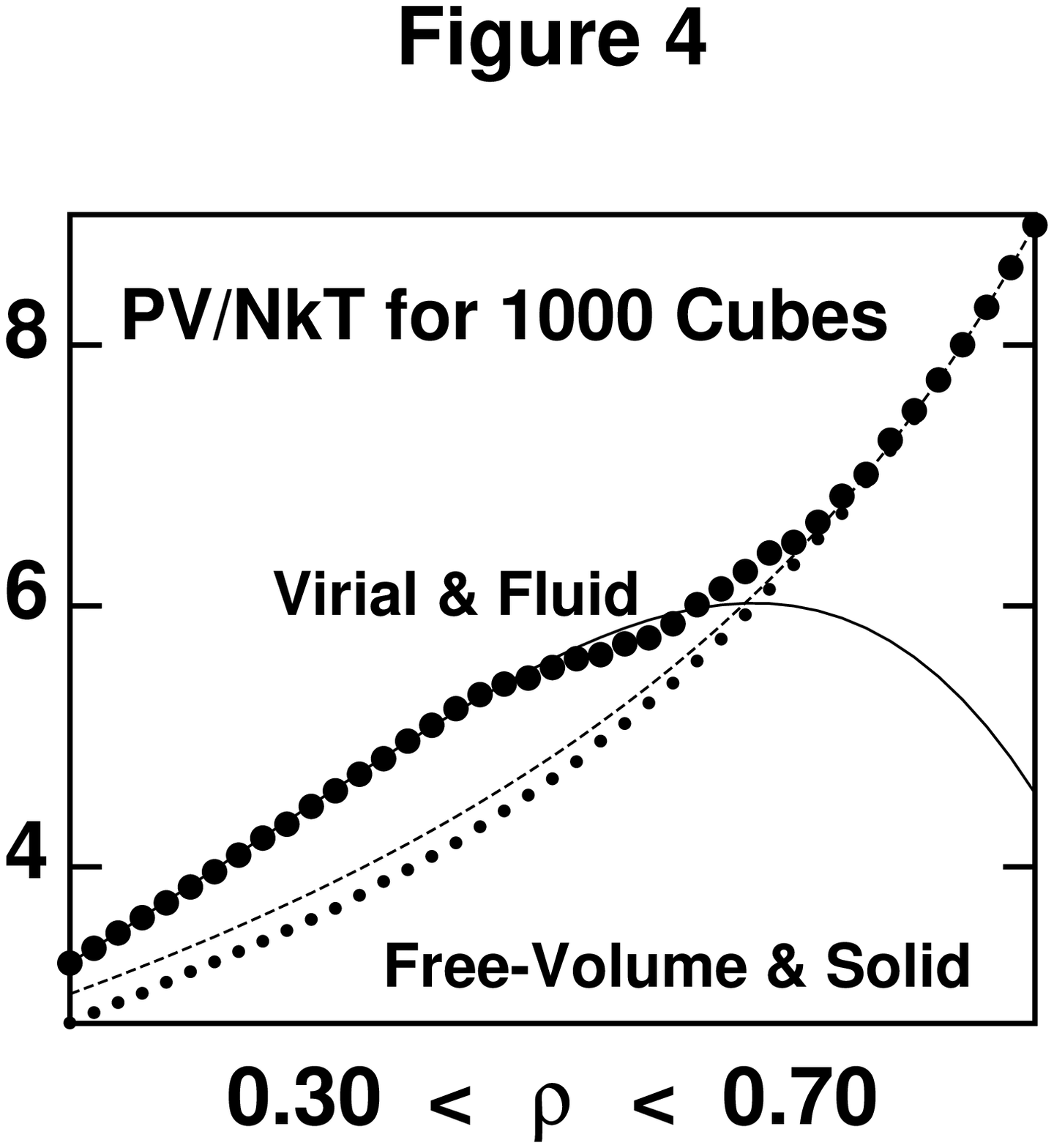}
\caption{
Compressibility factor for 1000 hard parallel cubes (dots) compared with
the predictions (lines) of the 7-term virial series and the
free-volume theory.  The upper set of dots represents unconstrained
molecular dynamics while the lower set represents single-occupancy
simulations.  Each point corresponds to a simulation with a million
collisions. The dots shown represent 41 simulations, equally
spaced in density from 0.30 to 0.70, inclusive.
}
\end{figure}

\subsection{Pressure Data}

To make contact with earlier work, and to provide data for thermodynamic
integration we have considered a wide range of densities for squares and
cubes.  Tables I and II compare a small sampling of the single-speed
molecular dynamics data of the types shown in Figures 3 and 4.  For both
squares and for cubes, these
data include both conventional and single-occupancy predictions, as well
as  the pressure and entropy predictions of the truncated virial series,
the Pad\'e approximant, and the self-consistent cell model.  Although we
have carried out a wide range of simulations, with density spacings of
0.01 or 0.005 and a wide range of system sizes, we list here only two
sets of data, sufficient that other workers could easily check the
consistency of their calculations with ours.  The tabulated data, as well
as those shown in the Figures, are quite representative of our body of
results, and have been chosen so that the reader can see the relative
usefulness of the various virial series and cell models to predicting and
interpreting the dynamical data.

The results we tabulate for squares (in the range $0.40 \le \rho \le 0.65$)
in Table I show that some of the higher virial coefficients from the Pad\'e
approximant are negative (because adding in the higher contributions
{\em reduces} the sum {\em below} that of the truncated series). In general,
for cubes as well as squares, the
truncated series are just as useful as are the Pad\'e approximants.  There
is a significant difference between the two approaches, truncated and
Pad\'e, beginning, for squares, at a density of about 0.70 and, for cubes,
at a density of about 0.50.  There are also enhanced fluctuations just
beyond these densities, so that the pressure data by themselves leave the
exact nature of the fluid-solid phase transition somewhat nebulous.
Despite this uncertainty, the present
data certainly show that the van der Waals loop found in the earlier
dynamics work\cite{b7} was an artefact of the short computer runs which
were possible in the early 1970s.

The results for cubes in Table II, and plotted in Figure 4 with many
additional points, do lead to one relatively straightforward conclusion:
for cubes there is no suggestion of a first-order phase transition.  The
jumpy nature of the cube equation of state for systems with less than
1000 particles disappears for longer runs and larger systems.  Even a
discontinuity in slope (second-order transition) looks doubtful for cubes.

In both two and three dimensions the free-volume equation of state is
evidently exact, within terms of order unity, near close
packing.  At the same time it is hard to predict with great confidence
precisely where the transition from fluid to solid is located or
what its order might be from pressure data alone.

  In an
attempt better to locate and characterize the square and cube
fluid-solid phase transitions we investigated the single-occupancy
entropy approach described in the following subsection.  This same
approach was successful forty years ago in interpreting hard-disk and
hard-sphere simulations\cite{b10}.

\subsection{Entropy Calculations and the Solid-Phase Entropy Constant}

Two thermodynamic phases with the same pressure, temperature, composition,
and Gibbs' free energy per particle,
$$
(G/NkT) = (E/NkT) + (PV/NkT) - (S/Nk) \ ,
$$
are in equilibrium with one another.  For squares and cubes the energies
of the fluid and solid are purely kinetic, $kT/2$ per degree of freedom,
so that the only difficulty in comparing free energies lies in estimating
the entropy $S$.  Ree and Hoover\cite{b10,b16,b17} showed how to implement
Kirkwood's single-occupancy thermodynamics\cite{b23} so as to measure
the entropy in the solid phase, $S_{\rm solid} \simeq S_{\rm SO}$.  The
cell-cluster theory is an alternative approach and was successful for
hard squares\cite{b6,b24}.  So far as we know this theory has not been
applied to parallel cubes until now.

The somewhat inconclusive nature of the pressure plots (Figures 3 and 4) led
us to consider separate calculations of the entropy for both phases,
fluid and solid.  Knowing the entropy is equivalent to knowing the
free energy for hard particles.  The fluid phase is no problem.  From
the virial series, the entropy, relative to that of an ideal gas at the
same density and temperature, can be expressed in terms of the virial
coefficients,
$$
(S/Nk) - (S/Nk)_{\rm ideal} = -B_2\rho - (B_3\rho^2/2) - (B_4\rho^3/3)
 - (B_5\rho^4/4) - \dots \ .
$$
The fluid-phase entropies for squares and cubes appear in Tables I 
and II.  The analytic virial-series entropy, Pad\'e approximant
entropy, and the entropy from integrated molecular dynamics pressures
are included there.

To calculate the isothermal solid-phase entropy we can use direct
integration of the single-occupancy equation of state:
$$
d(S/Nk)_T = -(PV/NkT)d\ln\rho \ .
$$
It is convenient to integrate the compressibility-factor difference,
$$
(\Delta S/Nk)_\rho = [(S/Nk)_{\rm SO} - (S/Nk)_{\rm FV}]_\rho =
\int_0^\rho [(PV/NkT)_{\rm FV} - (PV/NkT)_{\rm SO}]d \ln \rho^\prime \ ,
$$
using the known low-density values as the initial condition at 
$\rho = 0.01$:
$$
[\rho \simeq 0] \longrightarrow 
[S_{\rm SO} = S_{MD} - Nk = S_{\rm FV}] \ .
$$
$$
S_{\rm SO} \rightarrow \{-1 - 2\rho^{3/2},-1 - 6\rho^{4/3}\} \
{\rm for \ \{squares,cubes\}} \ .
$$
These limiting cases result if the lowest-order term in a Mayer
$f$-function expansion of the single-occupancy partition function
is worked out\cite{b25}.  Apart from the factor $-\rho /V$ this pair
interaction term corresponds to the product of (i) the number
of shared nearest-neighbor cell walls ($2N$ for squares and $3N$
for cubes) and (ii) the two-particle integral in the vicinity of
such a wall, [$2\rho ^{1/2}/2$ for squares and $4\rho ^{2/3}/2$ for
cubes].  Such a calculation was detailed for hard disks and spheres
in 1967\cite{b25}.
Because the single-occupancy pressure data are smooth and regular,
without large fluctuations, the numerical integrations are relatively
easy to perform, for both squares and cubes.  With a few dozen points the
trapezoidal rule can easily achieve an accuracy of $\pm 0.01Nk$.

Straightforward numerical integration of the single-occupancy data,
using the thermodynamic relation,
$$
\Delta S/Nk = \int -PV/NkT d\ln \rho \ ,
$$
shows that the entropy for hard squares, at densities of 0.82 and above,
exceeds that of the Eyring-Hirschfelder cell model by
$s_0({\rm squares}) = 0.27_3Nk$, in
precise agreement with the Rees' calculation\cite{b6} as well as the
corresponding result for hard disks\cite{b17}.  The last row of data
in Table I give the estimate (at $\rho = 0.80$),
$$
[S_{SO} - S_{EH}]/Nk = s_0({\rm squares}) =
5.497 - \ln(4) - 3.842 = 0.27 \ .
$$

The hard-cube entropy constant is somewhat less than that for hard
spheres\cite{b17}.  For cubes, with
$$
\Delta (PV/NkT) \equiv (PV/NkT)_{\rm Cell} - (PV/NkT)_{\rm SO} \ ,
$$
integration into the stable solid phase gives the
entropy constant as follows:
$$
s_0({\rm cubes}) = (S/Nk)_{\rm SO} - [(S/Nk)_{\rm FV} + \ln(8)] =
$$
$$
(S/Nk)_{\rm SO} -  (S/Nk)_{\rm EH} =
2.21 - 2.08 = 0.13 \ .
$$
Similarly, the last line of Table II, corresponding to $\rho = 0.70$,
gives:
$$
[S_{SO} - S_{EH}]/Nk = 7.565 - \ln(8) - 5.346 = 0.13 \  .
$$
The hard-cube configurational integral near close packing exceeds
that of the Eyring-Hirschfelder cell theory by a factor of
$e^{0.13} = 1.14$.  For hard spheres the corresponding factor is
$e^{0.216} = 1.24$.  In the following Section we consider the
usefulness of these entropy estimates in locating phase equilibria
for squares and cubes.

\section{Entropy and the Melting Transitions for Squares and Cubes}

Entropy plays a key role in establishing the nature of the melting
transition for squares and cubes.  At a fixed density the
hard-particle phase having the greater entropy has also the lesser
Helmholtz' free energy, $A = E - TS$, and is the stable phase.  Thus
the relative stability of the fluid and the solid is determined by
their relative entropies.

The difference in entropy between the stable fluid and the less-stable
single-occupancy solid was called the ``communal entropy'' by
Kirkwood\cite{b23}.  The communal entropy, absent in the 
single-occupancy solid, would be restored if multiple occupancy of all
the cells were allowed.  Notice that Tonks' exact calculation of the
``hard-rod'' partition function\cite{b21}, mentioned
in the Introduction, correctly accounts for multiple occupancy in the
simplest one-dimensional case.  

The communal entropy difference, fluid minus
single-occupancy solid, is equal to $Nk$ in the low density limit.  The
communal entropy gets smaller as the melting transition is approached,
and finally vanishes at the density of the melting solid.  In addition
to this number-independent effect there is an $N$-dependent contribution
$\Delta S_{\rm com} = k\ln N/N$ which can be ascribed to 
fluctuations\cite{b26}.  As a result, the fluid gains in stability as $N$
increases, so that the melting transition tends to higher pressures and
densities with increasing $N$.

Figures 5 and 6 show the communal entropy for squares and cubes based
on trapezoidal rule integration of the fluid and single-occupancy
solid data.  The hard-square data match, nearly perfectly, the
expected vanishing of the communal entropy (and equivalence of the
Helmholtz free energies) at the phase transition
density, 0.79. At that density both the entropies and the pressures
of the two phases, fluid and single-occupancy solid, are nearly equal.
Because the second-derivative isothermal bulk moduli,
$$
B_T = -V(\partial P/\partial V)_T = V(\partial^2A/\partial V^2)_T \ ,
$$
differ such a transition is called ``second-order'' rather than first.
Of course numerical work cannot distinguish between such a second-order
transition and a very weak first-order one, with slight differences in
the densities of coexisting phases.  The numerical work does make it
clear that the difference between the solid and fluid densities, if any, is
less than 0.01, considerably smaller than the corresponding 
solid-fluid density difference for hard disks\cite{b10,b25}.

Figure 5 illustrates the variation of communal entropy with density for 400
hard squares, shown as points, together with the predictions of the
truncated virial series through $B_7$ (line with a minimum at
$\rho = 0.80$) and those of the Pad\'e approximant (the lower line).
The number dependence seen in Table III can be avoided now by
simulating systems of thousands of particles for millions of collisions.
Such simulations are quite feasible on desktop computers.

\begin{figure}
\includegraphics[height=9cm,width=9cm,angle=-90]{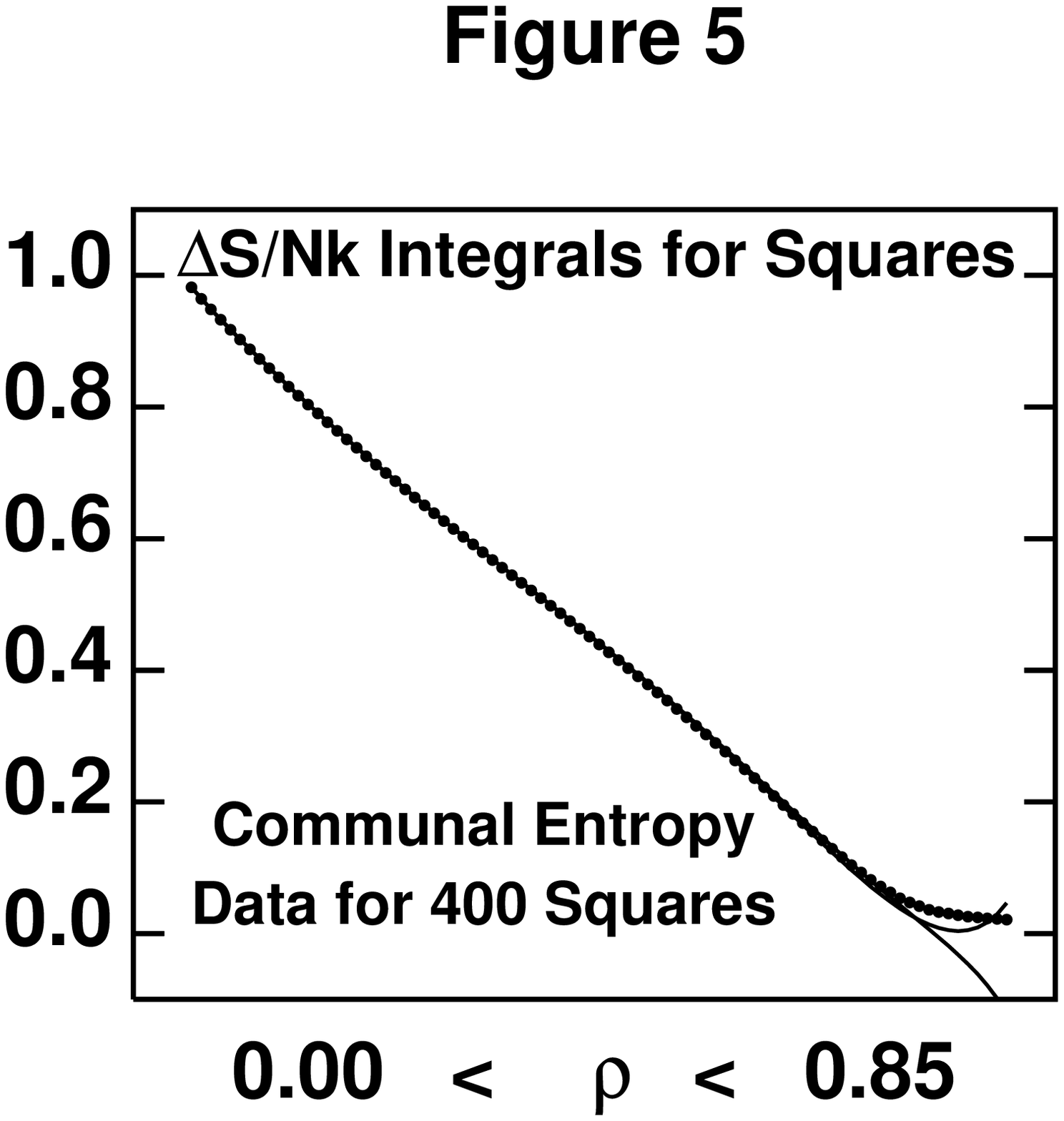}
\caption{
Entropy differences calculated by integration of the dynamic data for
400 hard parallel squares.  The points represent the ``communal
entropy'', the difference between the fluid and single-occupancy solid
entropies.  The upper line, with a minimum at $\rho = 0.80$ represents
the entropy from the virial series through $B_7$.  The lower line is
based on the hard-square Pad\'e approximant given in the Appendix.
The 85 fluid and solid simulations used to construct the entropy
differences used 400 000 and 4 000 000 collisions, respectively.
}
\end{figure}

\begin{figure}
\includegraphics[height=18cm,width=10cm,angle=-90]{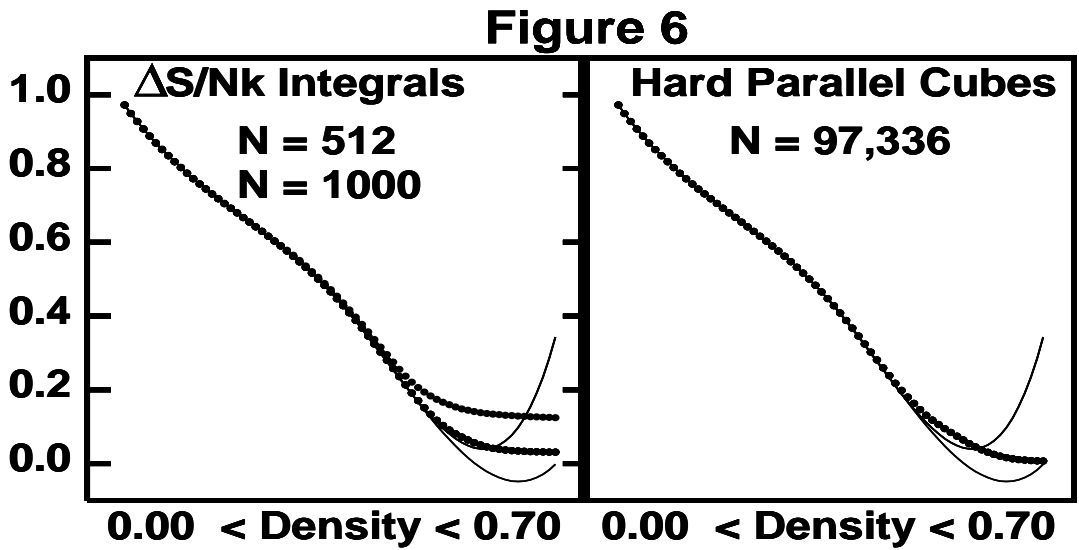}
\caption{
Entropy differences calculated by integration of the hard-cube
dynamic data.  The
points represent the difference between the fluid and solid entropies.
The lines represents the predictions of the truncated virial series
(above) and the Pad\'e approximant given in the Appendix (below).
The points for 512 and 1000 cubes were calculated from 70 simulations using
$1000N$ collisions.  The 97,336-particle data, with nearly a billion
collisions per point, are fully consistent with the single-occupancy
simulations with an entropy difference of less than $0.01Nk$  at the
maximum density shown here, $\rho = 0.70$.
}
\end{figure}

For cubes the number dependence complicates an analysis.  Systems with
no more than 512 particles exhibit an irregular behavior in the region
of confusion near the center of Figure 4. The unconstrained data for
1000 cubes, shown in Figure 6 and abstracted in Table II, are not quite
consistent with the single-occupancy calculations.  The high-density
entropy discrepancy is about 0.03$Nk$.

We took advantage of the University of Manchester cluster of processors
to complete an accurate unconstrained isotherm for $46 \times 46 \times
46 = 97,336$ hard cubes.  More data from that machine will be
forthcoming\cite{b27}.  The corresponding entropy data are shown in
Figure 6.  Figure 7 displays the difference between the 1000-cube and
97,336-cube compressibility factors.

\begin{figure}
\includegraphics[height=7cm,width=7cm,angle=-90]{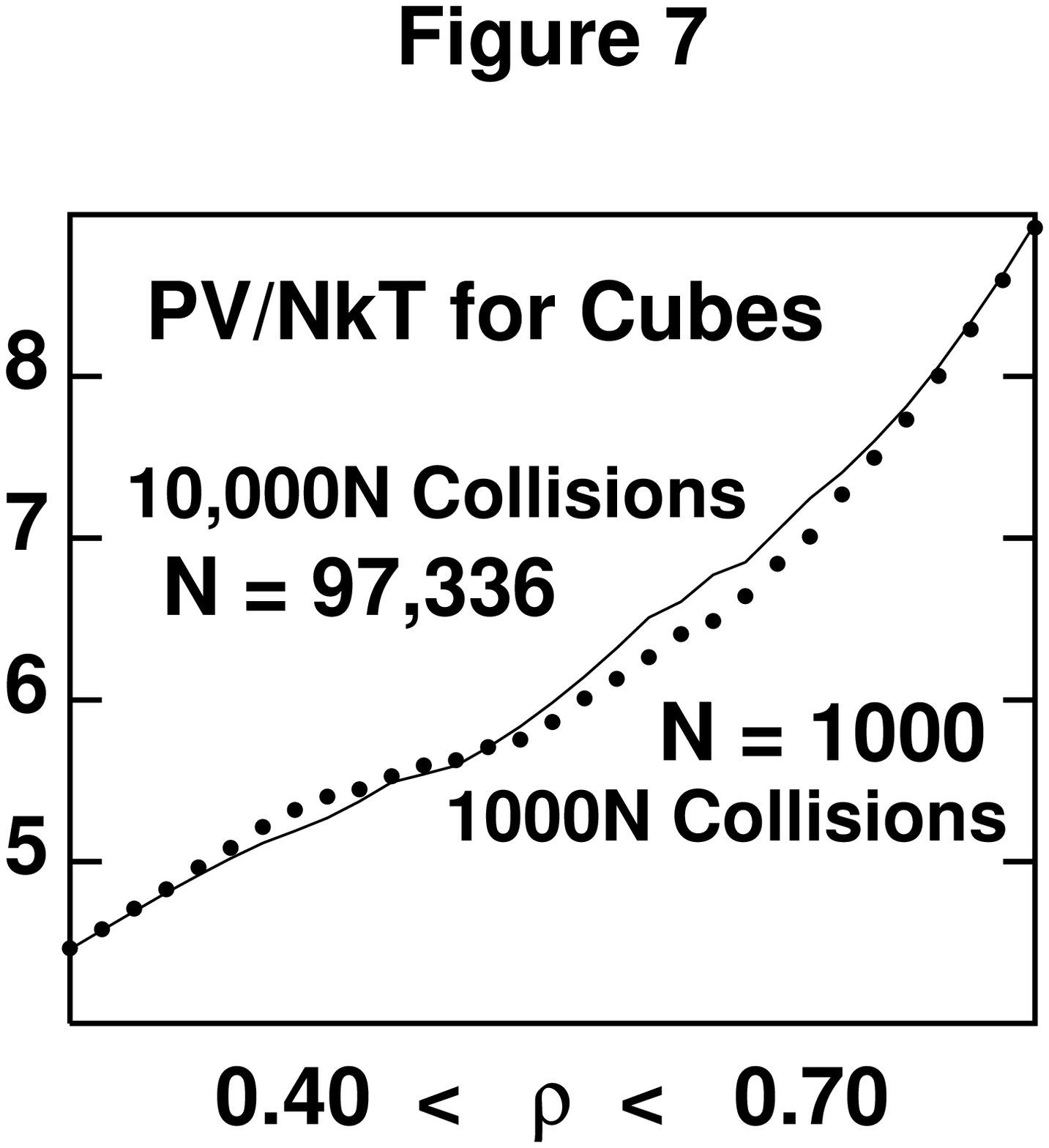}
\caption{
Number-dependence of the pressure.  31 simulations with $1000N$ collisions of
1000 cubes are compared with those with $10,000N$ collisions of 97,336 cubes.
}
\end{figure}

The interpretation of the relatively-smooth data for squares is more
straightforward. See Figure 8.  The Rees\cite{b6} reached the
conclusion that squares have no first-order phase transition and the
lack of difference between the ``fluid'' and ``solid'' equations of
state near $\rho = 0.79$ is quite consistent with this point of view.

\begin{figure}
\includegraphics[height=13cm,width=7cm,angle=-90]{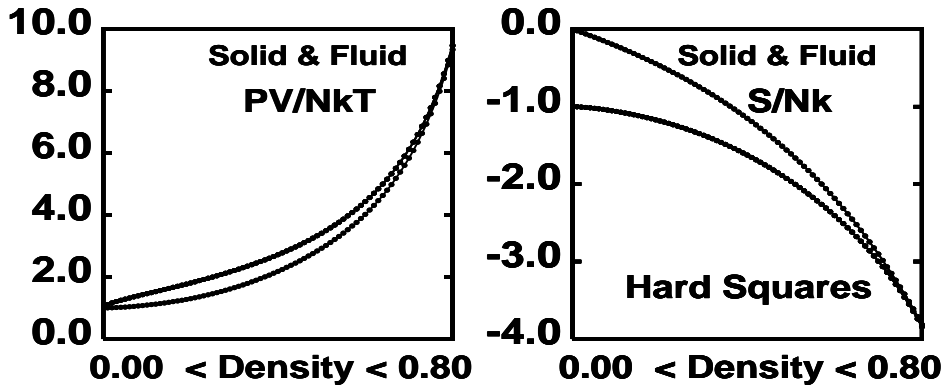}
\caption{
Single-speed molecular dynamics pressure and integrated entropy for
400 fluid squares and 900 single-occupancy solid squares using a density
interval of 0.01.
}
\end{figure}

Cubes exhibit much more hysteresis and number dependence than do squares. 
Figures 9 and 10 show the relatively slow convergence of the pressure for
densities in the region of confusion.  $10N$ collisions are scarcely
enough to distinguish the pressure from the free volume theory.  Longer
runs, with $10^4N$ collisions, show that with increasing time the pressure
gradually rises to a level between the truncated seven-term series
and the somewhat higher-pressure Pad\'e approximant. We have included some
longer-run data, for both squares and cubes, in Tables III and IV.

\begin{figure}
\includegraphics[height=13cm,width=7cm,angle=-90]{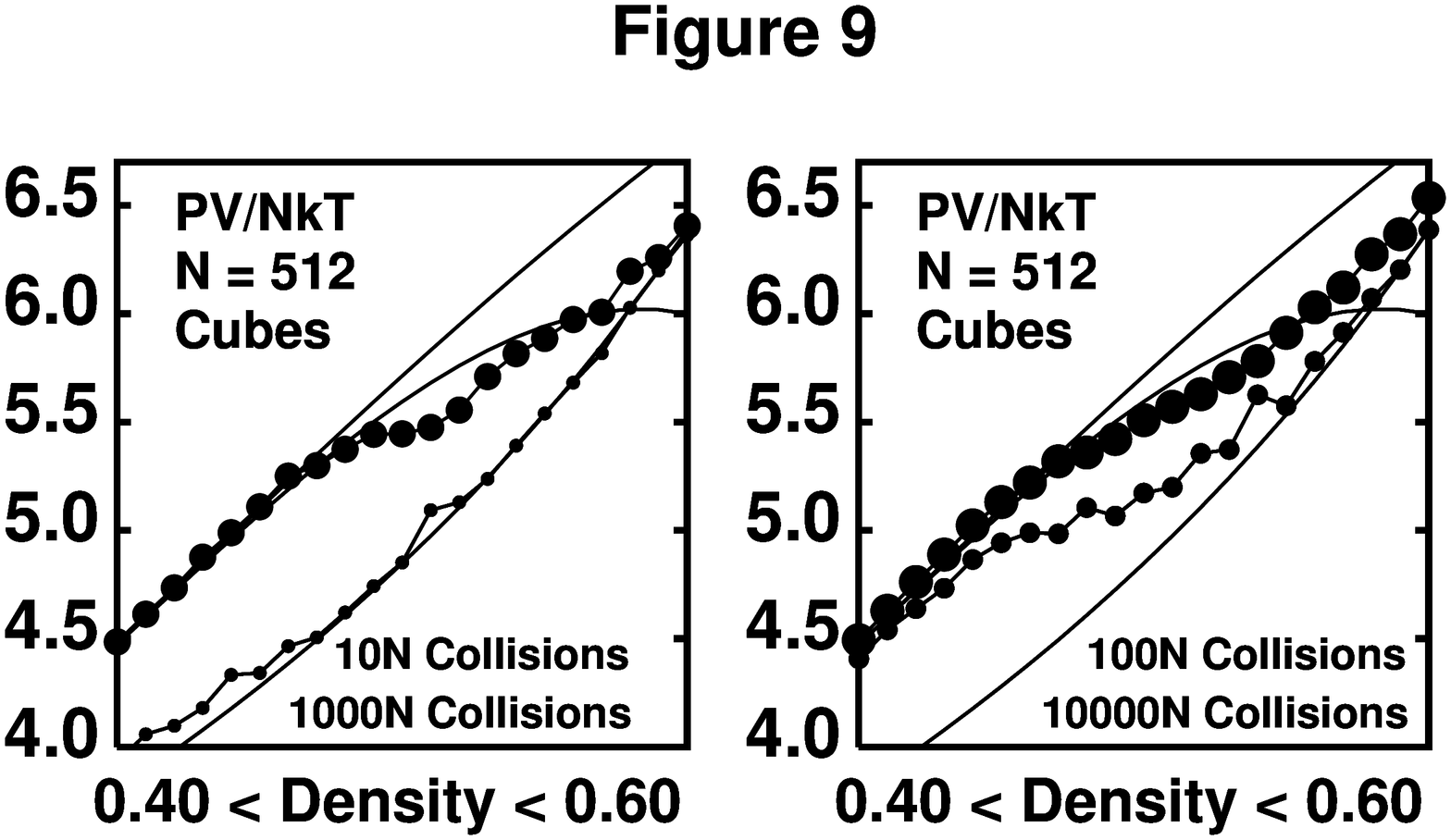}
\caption{
Pressure data for 512 cubes in the region of confusion.  Run lengths
of $10N$, $100N$, $1000N$, and $10000N$ collisions are indicated
with four increasing dot sizes and a density interval of 0.01.  The
curves are (from top to bottom) the Pad\'e approximant, the seven-term
virial series, and the free volume theory.
}
\end{figure}

\begin{figure}
\includegraphics[height=13cm,width=7cm,angle=-90]{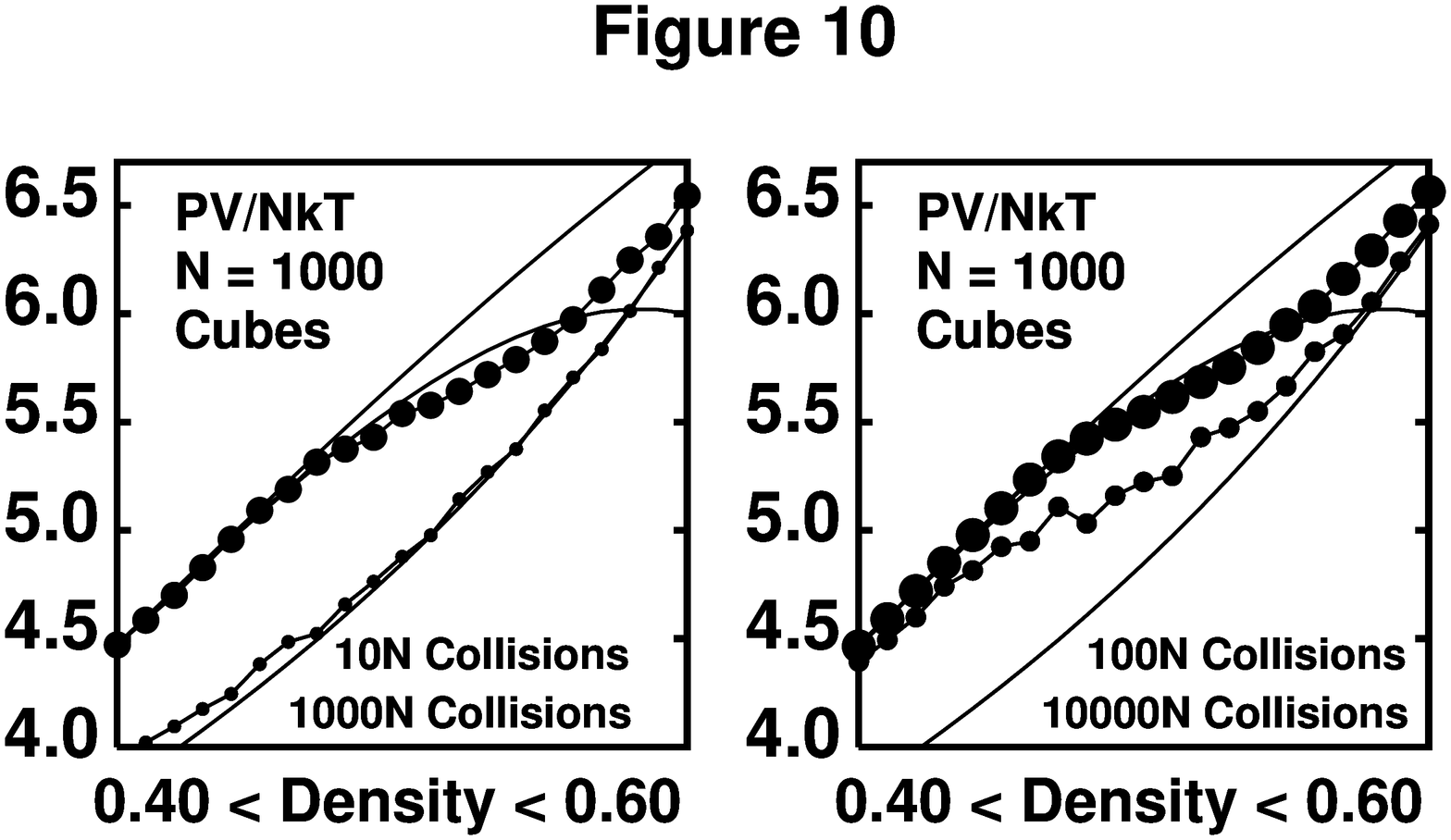}
\caption{
Pressure data for 1000 cubes in the region of confusion.  Run lengths
of $10N$, $100N$, $1000N$, and $10000N$ collisions are indicated
with four increasing dot sizes.  The curves are (from top to bottom)
the Pad\'e approximant, the seven-term virial series, and the free
volume theory.
}
\end{figure}

We also measured an ``irregular'' (fluid $\rightarrow $ glassy) isotherm
for hard cubes.  We placed  $(N < 11 \times 11 \times 11)$ particles
{\em randomly} on a regular array of $12 \times 12 \times 12 =1728$ lattice
sites in a volume $V=1728$.  Thus the initial state was a perfect lattice
with many vacancies.  Some of the resulting pressures are shown in Figure 11,
compared there with the 1000-particle isotherm, the seven-term virial series,
and the free-voume theory. It is evident that at densities of 0.57 and
above the irregular isotherm deviates substantially from that of a
``magic-number'' system selected to ``fit'' the periodic boundaries perfectly.

\begin{figure}
\includegraphics[height=7cm,width=7cm,angle=-90]{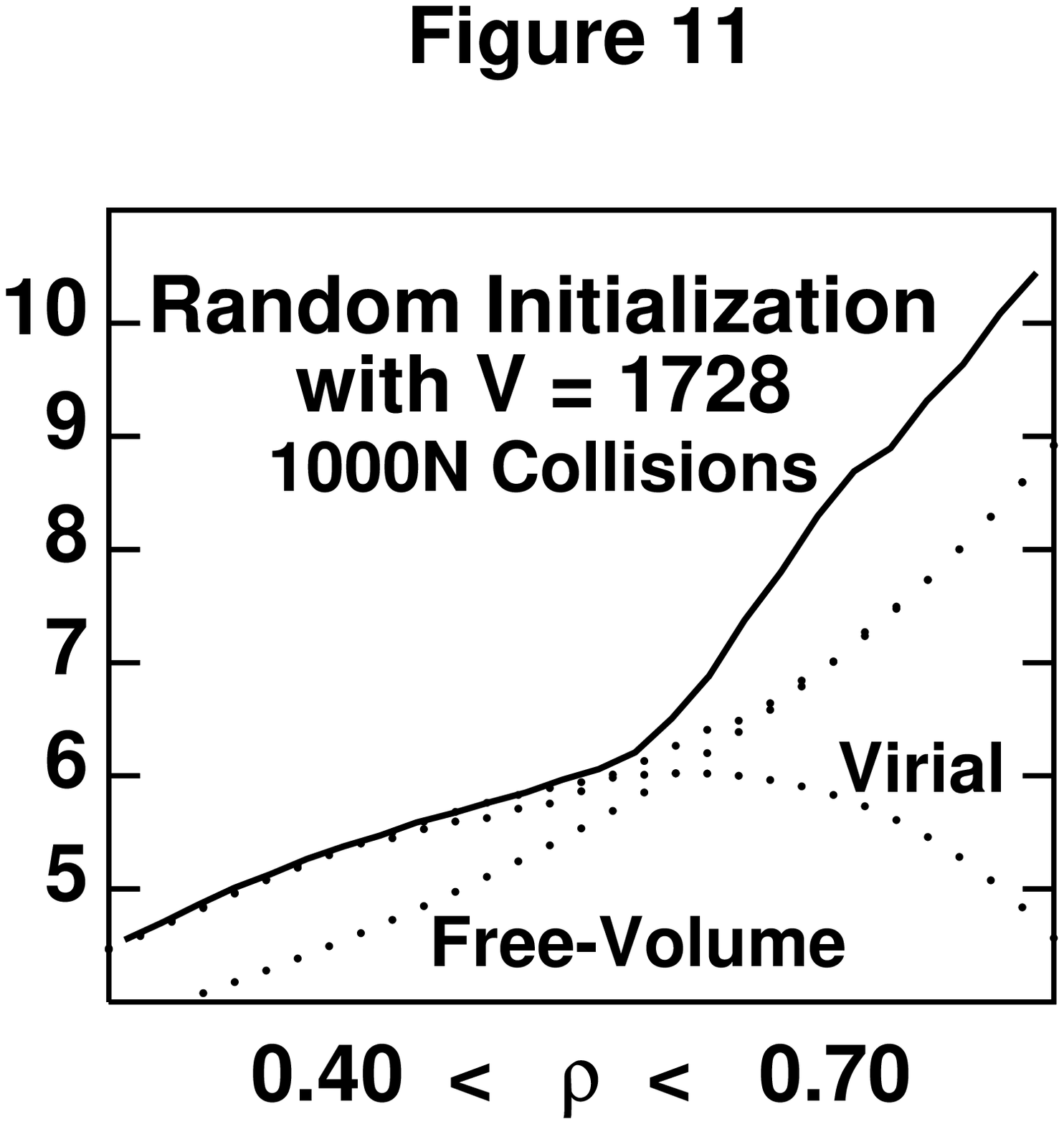}
\caption{
Pressure data for $N = 700, 720, 740 \dots$ 1200 cubes in a volume 1728.
Each simulation includes $1000N$ collisions.
Initial positions were chosen randomly from an $11 \times 11 \times
11$ lattice fitting the volume.  The results from the 26 simulations
are joined by a heavy line.  The dashed lines show the 1000-cube
isotherm, the seven-term virial series, and the free-volume theory
in the region of confusion. 
}
\end{figure}

The communal entropy for squares is relatively easy to compute.  Even
400 squares are sufficient to give a smooth equation of state with a
communal entropy close to zero at a density of 0.80.  See Figure 5.
To check this conclusion we have studied the hard-square density
region ($0.60 \le \rho \le 0.80$) carefully, with systems of 100,
400, 900, and 1600 particles, using simulations of at least one
million collisions.  Some results are summarized in Table III and
plotted in Figure 8.  The missing entropy in the integrated dynamic
pressure is about an order of magnitude smaller for squares than
for cubes, of the order $0.015Nk$ rather than $0.15Nk$.  The
calculated free energies for the fluid and solid phases merge very
smoothly at a density of about 0.793 so that there is no sharp phase
transition in the two-dimensional case.  In order to make a 
reproducible estimate for the transition location we represent the
hard-square fluid with the truncated virial series and the hard-square
solid by the free-volume equation of state.  The two pressures are equal
at $\rho = 0.793$:
$$
P_{virial}(0.793) = P_{fv}(0.793) = \rho kTZ = 7.242kT \ .
$$
and the entropy difference agrees precisely with the Rees' estimate 
and the hard-disk value:
$$
[S_{virial}(0.793) - S_{ideal}(0.793)]/Nk = -3.765 \ ; 
$$
$$
[S_{fv}(0.793) - S_{ideal}(0.793)]/Nk = -5.424 \longrightarrow
$$
$$
[S_{virial}(0.793) - S_{EH}(0.793)]/Nk = 0.27_3 \ ,
$$
where the Eyring-Hirschfelder cell-model entropy exceeds that of the
self-consistent free-volume theory for squares by $Nk\ln4 = 1.386Nk$.
The definite change in
slope required for consistency with the entropy data corresponds to
a second-order phase change, with no volume difference between the
two coexisting phases.

The detailed nature of the transition in the three-dimensional case
awaits larger-scale simulations or additional diagnostics.  The many
curvature changes in the data shown here could easily mask one or
more transitions of greater than second order.

\section{Summary and Conclusions}

Although computers are much faster now than in the pioneering days of
Alder, Jacobsen, Wainwright, and Wood\cite{b11,b12} the hard-cube
problem remains
a challenge.  Both the hard-square and hard-cube phase transitions are
weaker than the corresponding transitions for hard disks and spheres.
The precise nature of the cube transition remains uncertain.  The
square transition appears
to be second-order, with the pressure continuous and the compressibility
discontinuous.  The truncated virial series suggests a second-order 
transition while the higher-pressure Pad\'e version of the fluid would
correspond to a first-order transition a bit weaker than that found for
hard disks.  Such virial/Pad\'e
extrapolations of the pressure data are  useful tools for analyzing the
results for either squares or cubes.

For hard cubes in the solid phase, the free-volume equation of state and
the ordinary cell model are excellent descriptions of both the pressure
and the entropy (apart from an additive constant) for hard cubes, just
as they were for squares.  The characterization and appearance of the
hard-cube solid phase could be sharpened by (1) an evaluation
of the shear moduli, $C_{44}$ and $[C_{11}-C_{12}]/2$ and (2) a study
of the dependence of the diffusion coefficient on density. Both projects
are research challenges.  Protocols for measuring $C_{44}$ and $D$ in
the solid phase require innovative boundary conditions.

Jagla\cite{b13} found
a first-order melting transition for freely-rotating cubes.  He also
studied the parallel-cube model using constant-pressure simulations, and
described a ``continuous'' melting transition at a density of
$0.48 \pm 0.02$. Groh and Mulder presented an evenhanded criticism of
Jagla's work\cite{b14}, based on their own more extensive constant-pressure
simulations.  Groh and Mulder found a transition density of
$0.53_3 \pm 0.01$. In their view too the melting of hard parallel cubes is
probably ``continuous''. The free-energy uncertainty in their work,
$0.2NkT$, exceeds ours by an order of magnitude.  This difference seems
quite large, in that single-occupancy simulations can easily achieve
an accuracy an order of magnitude smaller, $0.01NkT$.  We are in
agreement with these two assessments of hard-cube melting as ``continuous''.

Beyond this exploration of the melting transitions, this work has some
interesting pedagogical consequences related to (i) the uncoupling of the
configurational and kinetic parts of the partition function and (ii) the
lack of coupling between the $x$ and $y$ and $z$ collisional momentum
changes.  The first of these uncouplings leads to successful but very
simple implementations of quasiergodic single-speed dynamics.   The
second uncoupling makes the hard-parallel-cube gas an ideal mechanical
thermometer, quite capable of measuring the independent tensor
components of the kinetic temperature\cite{b28}.  The simple linear
trajectories of the present model can also be generalized to
continuous potentials by using Lagrange multipliers to conserve energy
along straightline trajectories:
$$
\ddot x = \Lambda \dot x \ ; \ \ddot y = \Lambda \dot y \ ; \
\ddot z = \Lambda \dot z \ .
$$

The Lyapunov instability of hard squares and cubes would also make an
interesting topic for investigation.  Although the collisions are between
flat surfaces, without exponential growth in a scattering angle, at the
same time it is clear that an offset in the particle coordinates will
eventually (in a time roughly proportional to the offset) lead to a
missed collision, with a totally different subsequent evolution.  To
relate these collisional bifurcations to standard Lyapunov analyses is
another challenging research goal.

\section{Acknowledgments}

We thank Les Woodcock, Francis Ree, and Leo Lue for encouragement,
useful comments, and a selection
of references.  This work was presented at the University of Manchester
in the spring of 2009.

\newpage

\section{Appendix}

Pad\'e approximants to the seven-term virial series for squares and
cubes can be obtained by equating the coefficients of like powers of
the density.  The symmetric approximants are
$$
PV/NkT = \frac{1 - 0.98155\rho + 0.32754\rho^2 - 0.02760\rho^3}
              {1 - 2.98155\rho + 3.29065\rho^2 - 1.33090\rho^3}
$$
for squares, and
$$
PV/NkT = \frac{1 + 1.45948\rho + 2.28842\rho^2 + 0.91523\rho^3}
              {1 - 2.54052\rho + 3.45049\rho^2 - 1.35540\rho^3}
$$
for cubes.  Some details of the computation are given in Reference 17.

\newpage
\noindent
Table I. Compressibility factor and reduced entropy (relative to an ideal
gas at the same density and temperature) for 400 hard parallel squares with
40,000 collisions at each density.  $s \equiv (S_\rho - S_{\rm ideal})/Nk$.

\vspace{.5in}

\begin{tabular}{| c || c || c || c || c || c || c || c || c || c || c |  }
        \hline
$\rho$&$Z_{MD}$&$Z_{virial}$&$Z_{Pad\acute{e}}$&$Z_{SO}$&$Z_{FV}$
      &$s_{MD}$&$s_{virial}$&$s_{Pad\acute{e}}$&$s_{SO}$&$s_{FV}$  \\
\hline

0.05&1.108&1.108& 1.108&1.032&1.288&-0.105 &-0.104& -0.104&-1.022 &-1.506 \\
0.10&1.231&1.234& 1.234&1.095&1.462&-0.217 &-0.216& -0.217&-1.062 &-1.760 \\
0.15&1.382&1.382& 1.382&1.184&1.632&-0.339 &-0.338& -0.339&-1.116 &-1.980 \\
0.20&1.551&1.556& 1.556&1.304&1.809&-0.472 &-0.471& -0.472&-1.185 &-2.186 \\
0.25&1.760&1.763& 1.763&1.459&2.000&-0.618 &-0.617& -0.617&-1.269 &-2.386 \\
0.30&2.005&2.008& 2.008&1.657&2.211&-0.778 &-0.777& -0.778&-1.369 &-2.587 \\
0.35&2.317&2.299& 2 299&1.897&2.449&-0.955 &-0.954& -0.954&-1.488 &-2.791 \\
0.40&2.647&2.648& 2.646&2.182&2.721&-1.151 &-1.149& -1.150&-1.626 &-3.002 \\
0.45&3.058&3.064& 3.059&2.524&3.038&-1.369 &-1.367& -1.367&-1.784 &-3.222 \\
0.50&3.541&3.561& 3.550&2.931&3.414&-1.611 &-1.609& -1.608&-1.965 &-3.456 \\
0.55&4.096&4.156& 4.135&3.419&3.870&-1.882 &-1.880& -1.878&-2.171 &-3.707 \\
0.60&4.782&4.867& 4.833&4.022&4.436&-2.184 &-2.184& -2.180&-2.407 &-3.980 \\
0.65&5.586&5.714& 5.678&4.805&5.161&-2.523 &-2.526& -2.519&-2.678 &-4.282 \\
0.70&6.349&6.724& 6.740&5.838&6.122&-2.903 &-2.912& -2.903&-2.996 &-4.624 \\
0.75&7.575&7.924& 8.181&7.268&7.464&-3.329 &-3.346& -3.345&-3.376 &-5.020 \\
0.80&9.476&9.346&10.426&9.339&9.472&-3.814 &-3.837& -3.874&-3.842 &-5.497 \\
\hline
\end{tabular}

\newpage
\noindent
Table II. Compressibility factor and reduced entropy (relative to an ideal
gas at the same density and temperature) for 1000 fluid hard parallel cubes
and 1000 single-occupancy hard parallel
cubes with 1,000,000 collisions at each density.  $s \equiv (S_\rho - S_{\rm ideal})/Nk$.

\vspace{.5in}

\begin{tabular}{| c || c || c || c || c || c || c || c || c || c || c |  }
        \hline
$\rho$&$Z_{MD}$&$Z_{virial}$&$Z_{Pad\acute{e}}$&$Z_{SO}$&$Z_{FV}$
      &$s_{MD}$&$s_{virial}$&$s_{Pad\acute{e}}$&$s_{SO}$&$s_{FV}$  \\
\hline

0.05& 1.224& 1.224& 1.224& 1.129& 1.583& -0.212& -0.212& -0.212& -1.100 & -2.379 \\
0.10& 1.501& 1.501& 1.501& 1.346& 1.866& -0.449& -0.449& -0.449& -1.252 & -2.872 \\
0.15& 1.840& 1.840& 1.840& 1.641& 2.134& -0.715& -0.715& -0.715& -1.446 & -3.274 \\
0.20& 2.246& 2.247& 2.247& 1.995& 2.408& -1.011& -1.011& -1.011& -1.678 & -3.637 \\
0.25& 2.725& 2.723& 2.721& 2.392& 2.702& -1.339& -1.339& -1.339& -1.942 & -3.982 \\
0.30& 3.261& 3.264& 3.260& 2.804& 3.025& -1.700& -1.700& -1.699& -2.232 & -4.321 \\
0.35& 3.845& 3.857& 3.850& 3.197& 3.387& -2.092& -2.093& -2.091& -2.539 & -4.660 \\
0.40& 4.464& 4.475& 4.472& 3.597& 3.799& -2.512& -2.515& -2.512& -2.858 & -5.005 \\
0.45& 5.085& 5.075& 5.101& 4.081& 4.279& -2.955& -2.959& -2.958& -3.191 & -5.361 \\
0.50& 5.528& 5.594& 5.713& 4.676& 4.847& -3.412& -3.416& -3.422& -3.545 & -5.735 \\
0.55& 5.863& 5.943& 6.288& 5.410& 5.535& -3.857& -3.872& -3.898& -3.929 & -6.133 \\
0.60& 6.648& 6.000& 6.815& 6.317& 6.387& -4.309& -4.307& -4.381& -4.351 & -6.563 \\
0.65& 7.496& 5.608& 7.292& 7.436& 7.476& -4.785& -4.695& -4.866& -4.818 & -7.035 \\
0.70& 8.920& 4.565& 7.725& 8.904& 8.921& -5.315& -5.003& -5.348& -5.346 & -7.565 \\
\hline
\end{tabular}

\newpage
\noindent
Table III. Compressibility factors for hard squares in the
vicinity of the melting transition. 10,000 collisions per particle for
$N=$ 100 and 400; 1000 collisions per particle for $N=$ 900 and 1600.

\vspace{.5in}

\begin{tabular}{| c || c || c || c || c || c || c || c || c |  }
        \hline
$\rho$&$Z_{100}$&$Z_{400}$&$Z_{900}$&$Z_{1600}$
      &$Z_{virial}$&$Z_{Pad\acute{e}}$&$Z_{FV}$  \\
\hline

0.65 & 5.724 & 5.655 & 5.637 & 5.630 & 5.714&  5.678& 5.161  \\
0.70 & 6.549 & 6.682 & 6.641 & 6.608 & 6.724&  6.740& 6.122  \\
0.75 & 7.535 & 7.822 & 7.827 & 7.892 & 7.924&  8.181& 7.464  \\
0.80 & 9.410 & 9.510 & 9.488 & 9.484 & 9.472& 10.426& 9.472  \\
\hline
\end{tabular}
 
\vspace{.5in}

\noindent
Table IV. Compressibility factors for unconstrained periodic hard cubes in
the vicinity of the melting transition. 10,000 collisions per particle.

\vspace{.5in}

\begin{tabular}{| c || c || c || c || c || c || c || c || c || c |  }
        \hline
$\rho$&$Z_{64}$&$Z_{216}$&$Z_{512}$&$Z_{1000}$&$Z_{1728}$
      &$Z_{virial}$&$Z_{Pad\acute{e}}$&$Z_{FV}$  \\
\hline
0.45 & 4.733 & 5.062 & 5.127 & 5.106 & 5.067 & 5.075 & 5.101 & 4.279  \\
0.50 & 4.887 & 5.377 & 5.509 & 5.547 & 5.599 & 5.594 & 5.713 & 4.847  \\
0.55 & 5.461 & 5.777 & 5.912 & 5.948 & 5.971 & 5.943 & 6.288 & 5.535  \\
0.60 & 6.264 & 6.386 & 6.534 & 6.563 & 6.602 & 6.000 & 6.815 & 6.387  \\
0.65 & 7.286 & 7.450 & 7.513 & 7.529 & 7.521 & 5.608 & 7.292 & 7.476  \\
0.70 & 8.650 & 8.868 & 8.913 & 8.920 & 8.921 & 4.565 & 7.725 & 8.921  \\
\hline
\end{tabular}

\end{document}